\documentclass[pra,twocolumn,showpacs,preprintnumbers,amsmath,amssymb,superscriptaddress]{revtex4-1}

\usepackage{graphicx}
\usepackage{dcolumn}
\usepackage{bm}
\usepackage{mathbbol}
\usepackage{color}
\usepackage[FIGTOPCAP,raggedright,nooneline]{subfigure}

\newcommand{\Bra}{\langle}
\newcommand{\Ket}{\rangle}

\newcommand{\ignore}[1]{}

\newcommand{\cnb}[1]{{\color{black} #1}}

\begin{document}


\title{Self-Consistent Projection Operator Theory in Nonlinear Quantum Optical Systems: \\
A case study on Degenerate Optical Parametric Oscillators}

\author{Peter Degenfeld-Schonburg}
\email{peter.degenfeld-schonburg@ph.tum.de}
\affiliation{Technische Universit{\"a}t M{\"u}nchen, Physik Department, James Franck Str., 85748 Garching, Germany}
\author{Carlos Navarrete--Benlloch}
\email{carlos.navarrete@mpq.mpg.de}
\affiliation{Max-Planck-Institut f\"ur Quantenoptik, Hans-Kopfermann-str. 1, 85748 Garching, Germany}
\author{Michael J. Hartmann}
\email{m.j.hartmann@hw.ac.uk}
\affiliation{Institute of Photonics and Quantum Sciences, Heriot-Watt University, Edinburgh, EH14 4AS, United Kingdom}

\date{\today}

\begin{abstract}
\cnb{Nonlinear quantum optical systems are of paramount relevance for modern quantum technologies, as well as for the study of dissipative phase transitions. Their nonlinear nature makes their theoretical study very challenging and hence they have always served as great motivation to develop new techniques for the analysis of open quantum systems. In this article we apply the recently developed} self-consistent projection operator theory to the degenerate optical parametric oscillator to exemplify its general applicability to quantum optical systems.
We show that this theory provides an efficient method to calculate the full quantum state of each mode with high degree of accuracy, even at the critical point. It is equally successful in describing both the stationary limit and the dynamics,
\cnb{including regions of the parameter space where the numerical integration of the full problem} is significantly less efficient. We further develop a \cnb{Gaussian} approach consistent with our theory, which yields \cnb{sensibly better results than the previous Gaussian methods developed for this system, most notably}
standard linearization techniques.
\end{abstract}

\maketitle

%
%
\section{Introduction}
Nonlinear optical systems play an important role in the field of optics both in classical \cite{Boyd2003, Staliunas2002} and in quantum \cite{WMbook,Scullybook,Meystre1991,Hartmann07,Brandao08} regimes. Quantum mechanical effects, in particular, which are not explainable by classical optics, have triggered substantial research, especially in connection to modern applications such as high-precission measurements \cite{Goda08,Vahlbruch05,Treps03,Treps02} and quantum information communication and processing \cite{Braunstein05,Weedbrock12}. Importantly, the nonlinear nature of these systems leads to non-Gaussian states, which typically precludes an analytic treatment and therefore requires elaborate theoretical approaches \cite{CarmichaelBook1,CarmichaelBook2}.

In a system where the dynamical degrees of freedom evolve on different time scales, approximate descriptions of reduced complexity may be found. For example, \emph{adiabatic elimination} techniques can be exploited to derive effective equations of motion \cite{Mori65,ZollerBook}. In this work, we apply the recently introduced \emph{self-consistent projection operator theory} \cite{Degenfeld14} to the \emph{degenerate optical parametric oscillator}, and exemplify how it generalizes adiabatic elimination approaches. This theory takes dynamical back-action between the degrees of freedom into account and therefore does not require any time-scale separation. We expect our method to be directly applicable to other nonlinear quantum optical \cnb{models such as those for nondegenerate or multi-mode parametric oscillation \cite{Reid88,Drummond90,Navarrete08,Navarrete09}, lasing \cite{WMbook,Scullybook,MandelWolf,Breuer07}, optomechanical parametric oscillation \cite{OMPOCarlos,Nori}, or the dissipative \emph{Dicke model} \cite{DallaTorre,Parkins,Ritsch}}.

Degenerate optical parametric oscillators (DOPOs) have been extensively studied in the past \cite{Meystre1991,CarmichaelBook2} and are one of the paradigm examples of a system subject to a driven and dissipative phase transition. \cnb{It is formulated as a bosonic problem with two modes, \emph{signal} and \emph{pump}, subject to dissipation and interacting nonlinearly}. In the adiabatic limit of a fast decaying pump mode, an effective master equation can be derived by means of standard projection operator approaches \cite{CarmichaelBook2} and due to its reduced complexity, the steady state can be found by solving the corresponding Fokker-Planck equations for the positive P distribution \cite{Wolinsky88,Drummond80}. Yet away from the adiabatic limit one has to resort to numerical simulations or perturbative treatments \cite{Gardiner80,Kinsler93,Kinsler95,Chaturvedi02,Chaturvedi99,Pope00}. Non-equilibrium many-body techniques such as the Keldysh formalism have also been employed to study steady-state properties \cite{FleischhauerKeldysh,Mertens93,Swain93}. \cnb{While the application of all these techniques has allowed to deepen our understanding of DOPOs and phase transitions in driven dissipative quantum systems enormously, it is important to note that they are naturally built to determine the evolution of observables, making the determination of the quantum state of the optical fields very challenging, if not impossible.}

\begin{figure}[t]
\includegraphics[width=\columnwidth]{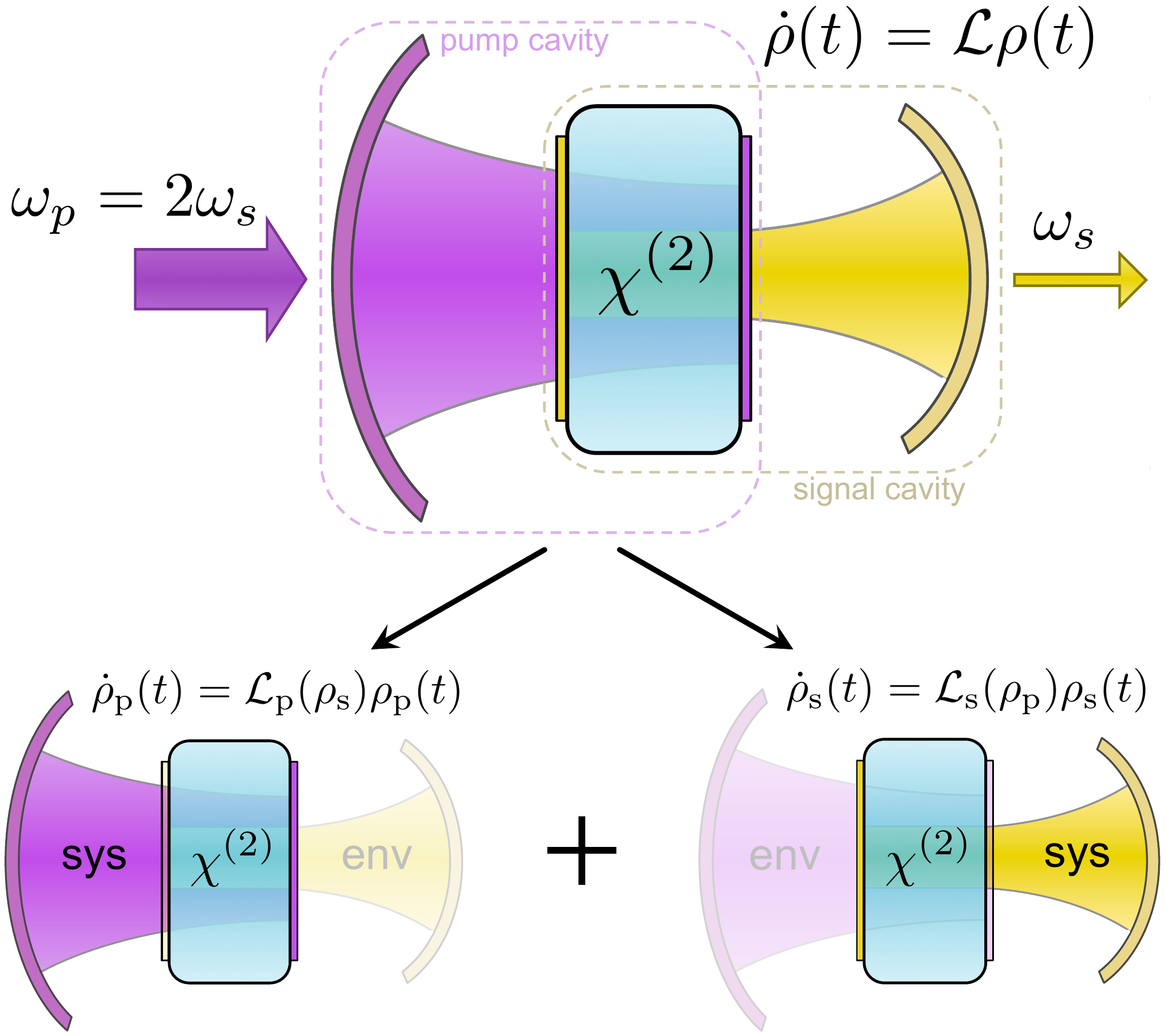}
\caption{\label{concept c-MoP} \cnb{Sketch of the self-consistent projection operator theory for the DOPO, which consists of an optical cavity containing a crystal with second-order optical nonlinearity, pumped by a laser at frequency $\omega_p=2\omega_s$ (pump mode), and capable of producing a field at the subharmonic frequency $\omega_s$ (signal mode) via down-conversion in the crystal. For the sake of illustration, we consider in the figure a doubly-resonant semi-monolithic configuration in which each face of the crystal acts as a mirror for one of the modes, but is transparent for the other, allowing to create independent cavities for the pump and signal modes via two additional partially transmitting mirrors \cite{Gigan06}. In our approach, the full problem} described by the state $\rho(t)$ and the Liouvillian $\mathcal L$ is mapped onto two coupled equations for the signal and pump modes. In one of the equations, the signal mode considered as the system is described by an effective master equation for its reduced state $\dot\rho_s(t)=\mathcal L_s(\rho_p)\rho_s(t)$ with an effective Liouvillian depending on the state of the pump, which plays here the role of an environment. The other equation considers the reversed scenario with the pump taking the role of the system while the signal is interpreted as the environment leading to the effective equation $\dot\rho_p(t)=\mathcal L_p(\rho_s)\rho_p(t)$. In this way the two equations form a closed set.}
\end{figure}

Our approach, in contrast, derives a set of coupled equations for the reduced states of the two \cnb{optical modes} of the DOPO. By numerically solving these equations, we find the reduced density matrices of both the pump and the signal modes. \cnb{We test the accuracy of our method by comparing its results with those of the full DOPO problem in regions of the parameter space where this is numerically tractable}. Our findings show that our method is remarkably close to the exact results, both for steady states and dynamics, while being less numerically demanding than the full simulation of the DOPO problem. It thus gives access to the reduced states of the cavity modes in regions of the parameters that are inaccessible to the latter.

The possibly largest reduction of complexity in nonlinear quantum optical systems, however, \cnb{comes from the application of Gaussian approximations} on the state of the system. Within a Gaussian theory one can \cnb{basically} cover the whole parameter space \cnb{efficiently to determine both steady-state and} dynamical quantities such as two-time correlation functions. The simplest and most widely used Gaussian approach is known as the \emph{linearization technique} \cnb{\cite{Drummond80,Lugiato81}, which consists in assuming that the system configuration is, on average, in its classical state, but is constantly driven out of it by some ``small'' quantum fluctuations. While this technique provides a good qualitative picture of the physics in many, albeit not all, systems, it leads to unphysical predictions close to the critical points of the classical theory, e.g., to infinite photon numbers in the case of the DOPO \cite{Collet84}. These unphysical predictions can be regularized by applying a more elaborate Gaussian state approximation where the system is not forced to stay in its classical state, but chooses instead an average configuration more consistent with the quantum fluctuations that perturb it \cite{Carlos14}}. Motivated by such an idea, we apply a \cnb{Gaussian approximation within the self-consistent projection operator theory}, and show that it gives more accurate quantitative results \cnb{than any of the usual Gaussian techniques}, as it does not assume a Gaussian state for the entire system, but only for the reduced state of one of the modes.

The remainder of the paper is organized as follows. In Sec.~\ref{Sec:theory} we \cnb{introduce the DOPO model}.  We also discuss its symmetries and briefly elaborate on the standard linearization approach in Sec.~\ref{sec:symmetries}. Sec.~\ref{c-MoP theory} reviews the main concepts of the self-consistent projection operator theory and introduces the \emph{self-consistent Mori projector} (c-MoP) equations, which lie at the center of our study. Our theory provides a systematic extension of mean-field approaches as demonstrated in Sec.~\ref{sec:MF} and reproduces known results in the \emph{adiabatic} and the \emph{diabatic} limits \cnb{introduced} in Sec.~\ref{sec:ada limit}. \cnb{An efficient procedure designed to deal with the \emph{non-Markovian} structure of the c-MoP equations is provided in Sec.~\ref{sec: born terms}, which we use in Sec.~\ref{sec: full numerics} to test} the accuracy of our method for steady-state quantities and to present \cnb{quantum states of the signal mode}. \cnb{A Gaussian state approximation on the c-MoP equations is performed in Sec.~\ref{sec:GSA}, which is shown to lead} to highly accurate quantitative results as compared to \cnb{previous} linearization techniques. \cnb{As a further test, we check in Sec.~\ref{sec:dyn} that our method provides the same level of accuracy for the dynamics, as it does for steady states}. Finally, we conclude our work and present an outlook in Sec.~\ref{conclusions}.

\section{The degenerate optical parametric oscillator}\label{Sec:theory}

\cnb{A DOPO consists of a driven optical cavity containing a crystal with second order optical nonlinearity, see Fig.~\ref{concept c-MoP}. Two relevant resonances at frequencies $\omega_s$ (\emph{signal} mode) and $\omega_p=2\omega_s$ (\emph{pump} mode) exist in the cavity, which are nonlinearly coupled via \emph{parametric down-conversion} inside the crystal, capable of transforming a pump photon into a pair of signal photons, and vice versa. We assume that the external driving laser is resonant with the pump mode.}
Including damping through the partially transmitting mirrors at rates $\gamma_p$ and $\gamma_s$ for the pump and signal modes, respectively, \cnb{the equation governing the evolution of the state $\rho$ of the system in a picture rotating at the laser frequency is given by \cite{Meystre1991,CarmichaelBook2}}
%
\begin{equation}\begin{split}\label{DOPO eqs}
\dot \rho(t)&= \left[\epsilon_p (a_p^\dagger-a_p)+\frac{\chi}{2} (a_p a_s^{\dagger \,2}-a_p^\dagger a_s^2)\, , \rho(t)\right] \\
&+\sum_{j=s,p} \gamma_j [2 a_j \rho(t) a_j^\dagger-a_j^\dagger a_j \rho(t)-\rho(t) a_j^\dagger a_j],
\end{split}
\end{equation}
\cnb{where $\chi/2$ is the down-conversion rate and $\epsilon_p$ is proportional to square root of the injected laser's power. We have defined bosonic operators $a_p$ and $a_s$ for the pump and signal modes, respectively, which satisfy canonical commutation relations $[a_j,a^\dagger_l]=\delta_{jl}$ and $[a_j,a_l]=0$. Note that the} nonlinear interaction is third order in the field operators, precluding a general analytic solution of Eq.~(\ref{DOPO eqs}) to which we refer as the \emph{Liouville-von Neumann equation} or simply the full master equation of the DOPO. 
\subsection{Linearization approach and symmetry breaking}\label{sec:symmetries}
The right hand side of Eq.~(\ref{DOPO eqs}) can also be written in a shorthand notation by introducing a \cnb{superoperator $\mathcal L$ (\emph{Liouvillian})}, such that $\dot \rho(t)=\mathcal L \rho(t)$. For the major part of this work, we will be interested in the steady
state $\rho_{ss}=\lim_{t\rightarrow \infty}\rho(t)$, which fullfills the equation $\mathcal L \rho_{ss}=0$. Due to the dissipation acting on both modes and because an arbitrarily large but finite truncation will always provide an arbitrarily good approximation, we expect the steady state to be unique \cite{Schirmer,Rivas12}. 

We further note the invariance of the Liouvillian under a unitary transformation \cnb{$U_2$ of Ising-type $Z_2$ which transforms $a_s$ as $U_2a_sU_2^\dagger=-a_s$.
Since the steady state is unique, this implies that it has to be invariant under the $Z_2$ transformation too, i.e. $U_2\rho_{ss}U_2^\dagger=\rho_{ss}$}. This in turn leads to vanishing steady state expectation values which include odd powers of the signal field operator $a_s$. In particular $\Bra a_s\Ket=0=\Bra a_p a_s^\dagger\Ket$, as for example \cnb{$\Bra a_s\Ket=\text{Tr}\{a_s \rho_{ss}\}=\text{Tr}\{U_2a_sU_2^\dagger U_2\rho_{ss}U_2^\dagger\}=-\Bra a_s\Ket$}.

However, the most \cnb{common technique used to analyze Eq.~(\ref{DOPO eqs}), known as the \emph{linearization approach}, breaks this $Z_2$ symmetry \cite{Drummond80,Lugiato81}, which has to be restored ``by hand'' at the end of the calculation, following the procedure that we explain at the end of Sec.~\ref{sec: full numerics}. Even though this method is more naturally introduced in the Heisenberg picture using the language of quantum Langevin equations, it also admits a Schr\"odinger picture interpretation in terms of two successive approximations in the master equation.
It starts by writing the bosonic operators as $a_j = \alpha_j+\delta a_j$, with $\alpha_j = \Bra a_j \Ket$ and hence $\Bra \delta a_j \Ket = 0$. In the first approximation, the fluctuation operators $\delta a_j$ are neglected altogether; the evolution equations for $\Bra a_j \Ket$ (Bloch equations), then provide a set of nonlinear differential equations for the amplitudes $\alpha_j$, which in the case of the DOPO read}
%
\begin{equation}\begin{split}\label{class eqs}
\dot{\alpha}_p=& \,\epsilon_p-\gamma_p \alpha_p-\frac{\chi}{2} \alpha_s^2 \\
\dot{\alpha}_s=&-\gamma_s \alpha_s+\chi \alpha_p \alpha_s^*.
\end{split}
\end{equation}
These correspond to the classical equations of the system, as they could have been obtained directly from Eq.~(\ref{DOPO eqs}) by assuming a coherent state for $\rho_{ss}$, \cnb{or simply from Maxwell's equations}. Depending on the \emph{injection parameter} $\sigma=\chi\epsilon_p/\gamma_s\gamma_p$ one finds two types of steady-state solutions of Eq.~(\ref{class eqs}). \cnb{One of them has $\alpha_s=0$ and $\alpha_p= \epsilon_p/\gamma_p$, and hence it does not break the symmetry; it is known as the \emph{below-threshold solution}, and is only stable for $\sigma < 1$. The other solution is bistable and has $\chi \alpha_s=\pm \sqrt{2 (\chi\epsilon_p-\gamma_s)}$ and $\chi \alpha_p=\gamma_s$, hence breaking the $Z_2$ symmetry; it is known as the \emph{above-threshold solution}, and exists only for $\sigma>1$.} The threshold point $\sigma=1$ marks a \emph{critical point} where the classical theory predicts a phase transition from a signal-off phase with $\alpha_s=0$ to a signal-on phase with $\alpha_s\neq0$. In the signal-off phase all injected power $\epsilon_p$ goes into the pump mode, while \cnb{after crossing the critical point all the extra injection is transferred to the signal mode}, see the gray thin solid line in Fig.~\ref{num c-MoP}.

\cnb{Once the classical solutions have been identified, the second approximation consists in coming back to the original master equation with the bosonic operators written as $a_j = \alpha_j+\delta a_j$, and neglecting any term which goes beyond quadratic order in the fluctuation operators $\delta a_j$. This leads to a so-called \emph{linearized master equation} which can be easily solved.}

\cnb{One has to keep in mind that this linearized theory can only be trustworthy when the classical solution is a strong attractor, because only then the quantum fluctuations driving the system out of equilibrium are strongly damped, and quantum noise can be treated as a small perturbation. This means that, in particular, any predictions obtained through this method cannot be trusted in the vicinities of critical points of the classical theory: points of the parameter space where one solution becomes unstable, making way for a new solution to kick in, hence creating non-analytic behaviour in some observable, that is, a classical \emph{phase transition}. Indeed, this is exactly the case for the DOPO, in which this linearized description breaks down at threshold, offering unphysical predictions such as infinite photon numbers in the signal field (as illustrated by the gray thin line in Fig.~\ref{GSA c-MoP}).}

%
%
\section{Self-consistent Mori Projector Approach}\label{c-MoP theory}
To explain the approach employed in our calculations, we will first recapitulate some basic ideas of the self-consistent projection operator theory \cite{Degenfeld14}. The first step is to divide the entire system into subsystems. In the DOPO \cnb{this naturally amounts to consider} the pump mode described by its reduced state $\rho_p(t)\equiv \text{Tr}_s \{\rho(t)\}$ and the signal mode described by $\rho_s(t)\equiv \text{Tr}_p \{\rho(t)\}$. In the spirit of \emph{open system theory} \cite{Zwanzig01,Breuer07} we will first treat the pump mode as an ``environment'' for the signal mode, which then takes the role of the open ``system''. Technically this is done by introducing the time-dependent, \emph{self-consistent Mori projector} $\mathcal{P}_t^p(\cdot)=\rho_p(t) \otimes \text{Tr}_p\{\cdot\}$ whose action on the full state $\rho(t)$ gives the factorized state $\mathcal{P}_t^p \rho(t)=\rho_p(t) \otimes \rho_s(t)$. The term ``self-consistent'' is chosen because the state of the pump in $\mathcal{P}_t^p$ is not a time-independent reference state but is rather obtained consistently from the time-evolving state $\rho(t)$ of the full dynamics. Using this projector, we derive a generalized \emph{Nakajima-Zwanzig} equation which is an exact equation for the reduced state of the signal mode \cite{Degenfeld14}. The effective Liouvillian describing such a Nakajima-Zwanzig equation will depend on the state of the pump $\rho_p(t)$. In order to obtain a closed set of equations we need to reverse the scenario and treat the pump mode as the ``system'' and the signal mode as the environment, see Fig.~\ref{concept c-MoP} for an illustration.

Again analogous to open system theory, we split the full Liouvillian $\mathcal L$ from Eq.~(\ref{DOPO eqs}) into three parts. After performing a displacement $a_p\rightarrow a_p+\tilde \alpha_p$, where $\tilde \alpha_p$ will be chosen later, see Sec.~\ref{sec:MF}, we write $\mathcal L= \mathcal L_p+\mathcal L_s+\mathcal L_I$, with
\begin{equation}\begin{split}\label{setup parts}
\mathcal L_p(\cdot)&= \left[ a_p^\dagger (\epsilon_p-\gamma_p \tilde \alpha_p)-a_p (\epsilon_p-\gamma_p \tilde \alpha_p^*)\,,\,\cdot\,\,\right]+\gamma_p D_{a_p}(\cdot) \\
\mathcal L_s(\cdot)&= \frac{\chi}{2} \left[ \tilde \alpha_p a_s^{\dagger \,2}-\tilde \alpha_p^* a_s^{2}\,,\,\cdot\,\,\right]+\gamma_s D_{a_s}(\cdot) \\
\mathcal L_I (\cdot)&= \frac{\chi}{2} \left[ a_p a_s^{\dagger \,2}-a_p^\dagger a_s^2\,,\,\cdot\,\,\right],
\end{split}
\end{equation}
where \cnb{we have defined the standard Lindblad superoperator $\mathcal D_{b}(\cdot)=2 b(\cdot)b^\dagger-b^\dagger b(\cdot)-(\cdot)b^\dagger b$, with $b$ being an arbitrary operator}. The displacement $a_p\rightarrow a_p+\tilde \alpha_p$ moves the large coherent background of the pump field into the free evolution of the signal $\mathcal L_s$, keeping only the pump mode's fluctuations within the nonlinear signal-pump interaction $\mathcal L_I$. Such a step is important as our theory expands in powers of the interaction Liouvillian $\mathcal L_I$ in order to solve the Nakajima-Zwanzig equation. As in reference \cite{Degenfeld14} we will expand to second order in the system-environment interaction. This approximation is known as the \emph{Born approximation} \cite{Breuer07}. 
The effective equations of the signal and the pump mode then read, \cnb{
%
\begin{align}
\dot{\rho}_s&(t)=\mathcal{L}_s\rho_s(t)+\frac{\chi}{2}\left[a_s^{\dagger 2}\Bra a_p\Ket(t)-a_s^2\Bra a_p\Ket\hspace{-0.06cm}^\ast\hspace{-0.06cm}(t),\rho_s(t)\right] 
\label{s c-MoP eqs}
\\
& +\left( \frac{\chi}{2}\right)^2\left\{\left[a_s^2\,,\int_{0}^{t}dt^\prime e^{\mathcal{L}_{s}(t-t^{\prime })}\mathcal{K}_s(t,t^\prime)\rho_s(t^\prime)\right]+\text{H.c.}\right\},  \nonumber
\end{align}
\begin{align}
\dot{\rho}_p&(t)=\mathcal{L}_p \rho_p(t)+\frac{\chi}{2}\left[a_p \Bra a_s^2\Ket\hspace{-0.06cm}^\ast\hspace{-0.06cm}(t)-a_p^\dagger\Bra a_s^2\Ket(t)\,,\rho_p(t)\right] 
\label{p c-MoP eqs} \\
& +\left(\frac{\chi}{2}\right)^2\left\{\left[a_p\,,\int_0^t dt^\prime e^{\mathcal{L}_p(t-t^\prime)}\mathcal{K}_p(t,t^\prime)\rho_p(t^\prime)\right]+\text{H.c.}\right\},  \nonumber
\end{align}
%
where we have defined the Kernel superoperators
\begin{align}\label{Ks}
\mathcal{K}_s(t,t^\prime)(\cdot)& =\delta a_s^2(t^\prime)(\cdot
)\,d_p^+(t,t^\prime)-(\cdot)\delta a_s^2(t^\prime)\,\tilde{d}_p^+(t,t^\prime)
\\
& -\delta a_s^{\dagger 2}(t^\prime)(\cdot)\,d_p^-(t,t^\prime)+(\cdot)\delta a_s^{\dagger 2}(t^\prime)\,\tilde{d}_p^-(t,t^\prime),  \nonumber
\end{align}
\begin{align}\label{Kp}
\mathcal{K}_p(t,t^\prime)(\cdot)& =\delta a_p(t^\prime)(\cdot)\,d_s^+(t,t^\prime)-(\cdot)\delta a_p(t^\prime)\,\tilde{d}_s^+(t,t^\prime)
\\
& -\delta a_p^\dagger(t^\prime)(\cdot)\,d_s^-(t,t^\prime)+(\cdot)\delta a_p^\dagger(t^\prime)\,\tilde{d}_s^-(t,t^\prime),\nonumber
\end{align}
and for any operator $A_j$ acting on the signal ($j=s$) or pump ($j=s$) subspace, we have defined the corresponding fluctuation operator $\delta A_j(t)\equiv A_j-\text{Tr}_j\{A_j\,\rho_j(t)\}$}.

The state of the pump mode $\rho_p(t)$ enters \cnb{the signal mode's dynamics, eq.~(\ref{s c-MoP eqs}),} via $\Bra a_p \Ket(t)\equiv\text{Tr}_p\{a_p\rho_p(t)\}$ and the correlation functions
\begin{align}\label{p corr func first}
 d_p^+(t,t')=\text{Tr}_p\{a_p^\dagger e^{\mathcal L_p(t-t')}\delta a_p^\dagger(t') \rho_p(t')\},\\
 \tilde d_p^+(t,t')=\text{Tr}_p\{a_p^\dagger e^{\mathcal L_p(t-t')} \rho_p(t') \delta a_p^\dagger(t')\},\nonumber\\
 d_p^-(t,t')=\text{Tr}_p\{a_p^\dagger e^{\mathcal L_p(t-t')}\delta a_p(t') \rho_p(t')\},\nonumber\\
 \tilde d_p^-(t,t')=\text{Tr}_p\{a_p^\dagger e^{\mathcal L_p(t-t')} \rho_p(t')\delta a_p(t')\}\nonumber,
\end{align}
In turn, the state of the signal mode $\rho_s(t)$ enters \cnb{the pump mode's dynamics, eq.~(\ref{p c-MoP eqs}),} via the expectation value $\Bra a_s^2 \Ket(t)\equiv\text{Tr}_s\{a_s^2\rho_s(t)\}$ and the correlation functions
\begin{align}\label{s corr func}
 d_s^+(t,t')=\text{Tr}_s\{a_s^{\dagger\,2} e^{\mathcal L_s(t-t')} \delta a_s^{\dagger\,2}(t') \rho_s(t')\},\\
 \tilde d_s^+(t,t')=\text{Tr}_s\{a_s^{\dagger\,2} e^{\mathcal L_s(t-t')} \rho_s(t') \delta a_s^{\dagger\,2}(t')\},\nonumber\\
 d_s^-(t,t')=\text{Tr}_s\{a_s^{\dagger\,2} e^{\mathcal L_s(t-t')} \delta a_s^2(t') \rho_s(t')\},\nonumber\\
 \tilde d_s^-(t,t')=\text{Tr}_s\{a_s^{\dagger\,2} e^{\mathcal L_s(t-t')} \rho_s(t') \delta a_s^2(t')\}\nonumber.
\end{align}

Equations~(\ref{s c-MoP eqs}) and~(\ref{p c-MoP eqs}) should be understood as two coupled equations which represent effective equations for the reduced states of the signal and the pump mode. We refer to these two equations as the \textbf{c-MoP} (\textbf{c}onsistent \textbf{Mo}ri \textbf{P}rojector) equations of the DOPO. They can be thought of as \emph{non-Markovian} and \emph{nonlinear master equations} which do not rely on any time-scale separation between the modes. We will elaborate in detail on the limits where time-scale separation is present in Sec.~\ref{sec:ada limit}. 

The only assumptions made so far are the Born approximation and the assumption of an initially factorized state $\rho(0)= \rho_p(0)\otimes\rho_s(0)$. The latter seems very reasonable by considering the vacuum as the state of the two modes before the driving laser is switched on. We also emphasize, 
our approach does not ignore system-environment or rather signal-pump correlations. In fact, it has been shown \cite{Degenfeld14} that the \emph{Born term}, the term second order in $\mathcal L_I$ which is here proportional to $(\chi/2)^2$, clearly takes signal-pump correlations into account. We will show the crucial importance of the Born term in several examples below. Of course, c-MoP theory or any theory based on the concept of projection operators does not give access to explicit expressions for system-environment correlation functions. An example in this context could be the cross-correlation function $\Bra a_p^\dagger a_s\Ket-\Bra a_p^\dagger \Ket \Bra a_s\Ket$. 

The most striking advantage of projection operator theories and in particular of the c-MoP theory is the reduction of the complexity of the problem. In the example of the DOPO the complexity of the Liouville-von Neumann eq.~(\ref{DOPO eqs}) scales as $\dim \mathcal H_s \times \dim \mathcal H_p$, where $\mathcal H_{s/p}$ denotes the Hilbert space of the signal/pump modes, while the complexity of the c-MoP equations scale as $\dim \mathcal H_s+\dim \mathcal H_p$. 
The self-consistent Mori-projector theory thus offers a very significant reduction of complexity.
%
%
%
%
\subsection{Mean-field Approximation}\label{sec:MF}
A merely approximate but very simple way of solving the c-MoP equations is to consider all terms up to first order in the interaction $\mathcal L_I$ only. Hence we drop all terms proportional to $\chi^2$ from eqs.~(\ref{s c-MoP eqs}) and~(\ref{p c-MoP eqs}). Within this approximation it does not make a difference whether the pump field is displaced or not. For simplicity we put the displacement $\tilde \alpha_p$ from eq.~(\ref{setup parts}) to zero and obtain two coupled equations
\begin{equation}
\begin{split}\label{MF eqs}
\dot \rho_p(t) &=  \left[ (\epsilon_p-\frac{\chi}{2} \Bra a_s^2\Ket^*)\,a_p^\dagger-\text{H.c.}\,,\rho_p(t)\right] +\gamma_p \mathcal D_{a_p} \rho_p(t),
\\
\dot \rho_s(t) &= \frac{\chi}{2} \left[ \Bra a_p\Ket a_s^{\dagger\,2}-\text{H.c.}\,,\rho_s(t)\right] +\gamma_s \mathcal D_{a_s} \rho_s(t) ,
\end{split}
\end{equation}
known as mean-field equations \cite{FleischhauerMF}. These equations are quadratic in the field operators and therefore it is straightforward to solve them either numerically for the dynamics or analytically for the fixed points \cite{FleischhauerMF,Carlos14}. The stationary state of the signal mode will be a Gaussian state \cite{Braunstein05,Weedbrock12,CarlosQI} centered around a vanishing field amplitude $\Bra a_s\Ket=0$ as the mean-field equations do not break the Ising-type $Z_2$ symmetry. The steady state of the pump mode will be a coherent state with an amplitude given by $\Bra a_p\Ket_{ss}^{MF}=(\epsilon_p-\frac{\chi}{2} \Bra a_s^2\Ket_{ss}^{MF})/\gamma_p$. 

Just like the c-MoP equations~(\ref{s c-MoP eqs}) and ~(\ref{p c-MoP eqs}), the mean-field equations are coupled nonlinear equations which have to be solved self-consistently. Within mean-field theory fluctuations of the pump mode are disregarded. Fluctuations of the signal mode, however, are (at least to some extend) taken into account \cite{FleischhauerMF,Carlos14}. This leads to the regularization of the divergences appearing in the classical theory or rather the standard linearization approach. For our purposes it is important to note that the pump field amplitude always stays below the classical \emph{above-threshold} solution, i.e. $\Bra a_p\Ket_{ss}^{MF} < \gamma_s/\chi$. In the remainder of the paper we will use it as the displacement in eq.~(\ref{setup parts}), i.e. $\tilde \alpha_p=\Bra a_p\Ket_{ss}^{MF}$. This will guarantee a well-behaved Liouvillian for the free system $\mathcal L_s$ as we will explain in more detail in Sec.\ref{sec: born terms}. 

The mean-field equations can also be found by putting the factorized state Ansatz $\rho(t)=\rho_p(t)\otimes\rho_s(t)$ into the Liouville-von Neumann equation, here given by eq.~(\ref{DOPO eqs}), before tracing out each of the modes separately. This well-known procedure, indeed, neglects all signal-pump correlations. Within the self-consistent projection operator theory, mean-field can be understood as an approximation to linear order in the interaction $\mathcal{L}_{I}$ for the dynamics of reduced density matrices. 
Our theory therefore provides a systematic generalization of mean-field approaches. It is due to the Born terms, which are second order in $\mathcal L_I$, that signal-pump correlations are taken into account. Therefore, we expect a different quality of approximation by going from first order to second order in the interaction.

\subsection{Born terms}\label{sec: born terms}

In order to solve the full c-MoP equations including the Born terms we will need to overcome two main difficulties. While the c-MoP equation~(\ref{p c-MoP eqs}) of the pump mode is quadratic in the field operators, granting us with a closed set of equations including only first and second moments of the pump field, the c-MoP equation~(\ref{s c-MoP eqs}) of the signal mode is quartic in the field operators. We will therefore either solve the equation of the signal fully numerically, see Sec.~\ref{sec: full numerics}, or apply a \emph{Gaussian state approximation} as presented in section Sec.~\ref{sec:GSA}. In any of these two approaches, we need to overcome the second difficulty which arrises due to the non-Markovian structure of our theory. In the remainder of this section we will thus show how to rewrite an integro-differential equation of first order into a set of coupled ordinary differential equations. For the present problem this step is crucial, as solving the integro-differential equations is significantly more demanding for both numerical and analytical approaches. 

 We start by evaluating the correlation functions of the pump. By taking derivatives of \cnb{the pump correlators $d^\pm_p(t,t')$ and $\tilde d^\pm_p(t,t')$ with respect to $t$}, see eq.~(\ref{p corr func first}), considering initial conditions at $t=t'$ (note that we understand from the c-MoP equations that $t'\leq t$), and exploiting the fact that the operator $\delta a_p^\dagger(t')\rho_p(t')$ is traceless, we find
\begin{equation}\label{p corr func}
\begin{split}
 d_p^+(t,t')&=\tilde d_p^+(t,t')=[\Bra a_p^{\dagger\,2}\Ket(t')-\Bra a_p\Ket^{\hspace{-0.06cm}* 2\hspace{-0.04cm}}(t')] e^{-\gamma_p (t-t')},
 \\
\tilde d_p^-(t,t')&=[1+\Bra a_p^{\dagger} a_p\Ket(t')-|\Bra a_p\Ket(t')|^2] e^{-\gamma_p (t-t')},
\\
  d_p^-(t,t')&=[\Bra a_p^{\dagger} a_p\Ket(t')-|\Bra a_p\Ket(t')|^2] e^{-\gamma_p (t-t')}.
 \end{split}
\end{equation}
Hence, all correlation functions of the pump can be written in a form where the $t$ dependence only enters in a simple exponential factor. 

A bit more effort is needed in order to simplify the correlation functions of the signal, but the main steps are mainly identical. All the functions in eq.~(\ref{s corr func}) are of the form $f(t,t')=\text{Tr}_s\{a_s^{\dagger\,2} e^{\mathcal L_s(t-t')} A(t')\}$ with a traceless operator $A(t')$ depending solely on $t'$. Again, we take the derivative of $f(t,t')$ with respect to $t$ and find an equation of motion of the form $\partial _t \vec{v}_{t'}(t)= M \vec{v}_{t'}(t)$ with a column vector
\begin{equation}
\vec{v}_{t'}(t)= \text{col}\left(\widetilde{\Bra a_s^\dagger a_s\Ket},\widetilde{\Bra a_s^2\Ket},\widetilde{\Bra a_s^{\dagger\,2}\Ket}\right),
\end{equation}
where the expectation values with the tilde are defined in the usual way as the trace over the signal mode but with a density matrix given by $\tilde\rho_{t'}(t)=e^{\mathcal L_s (t-t')} A(t')$. The matrix $M$ reads
\[
  M=
  \left( {\begin{array}{ccc}
   -2 \gamma_s & \chi \tilde\alpha_p & \chi \tilde\alpha_p^* \\
   2 \chi \tilde\alpha_p^* & -2 \gamma_s & 0\\
   2 \chi \tilde\alpha_p & 0 & -2 \gamma_s\\
  \end{array} } \right).
\]

It is straight forward to diagonalize $M$. \cnb{We write $M= U \Lambda U^{-1}$, with a similarity matrix $U$ that can be found analytically (but its expression is too lengthy to be reported here), and $\Lambda$ is the diagonal form of $M$ containing its eigenvalues $\lambda_1=-2 \gamma_s$, and $\lambda_{2,3}= -2\gamma_s\mp 2\chi |\tilde\alpha_p|$. We now solve for the vector $\vec{v}_{t'}(t)$, to find
\begin{equation}
\vec{v}_{t'}(t)=U e^{\Lambda (t-t')} U^{-1} \vec{v}_{t'}(t')\equiv \sum_{n=1}^3 M_n e^{\lambda_n (t-t')}\, \vec{u}_{A(t')}, \label{v_sol}
\end{equation} 
where we have defined the initial condition vector
\begin{equation}\label{signal cf vec}
 \vec{u}_{A(t')}=\vec{v}_{t'}(t')=\left(\begin{array}{c}
 \text{Tr}_s\{a_s^\dagger a_s A(t')\} \\ \text{Tr}_s\{a_s^2 A(t')\} \\ \text{Tr}_s\{a_s^{\dagger \,2} A(t')\} \\
 \end{array}\right),
\end{equation}
and the matrices $M_n = U \Pi_n U^{-1}$, where $\Pi_n$ is a projector in the $n$'th ``direction'', that is, a matrix with zeros everywhere but in element $(n,n)$ which is one.

Note that for the limit $\lim_{t\rightarrow \infty} \vec v _{t'}(t)$ to be uniquely defined, and therefore for $\mathcal L_s$ to be well-behaved, all the eigenvalues of $M$ must satisfy $\mathrm{Re}\{\lambda_n\} < 0$, which in turn leads to a requirement for the displacement $\chi \tilde \alpha_p < \gamma_s$}. This requirement is \cnb{indeed} fulfilled by choosing the mean-field displacement as mentioned above, see Sec.~\ref{sec:MF}. \cnb{In contrast, taking the classical solution as the displacement would lead to an ill-behaved $\mathcal L_s $ above and at the classical threshold point, that is, for $\sigma\geq1$.}    

Coming back to the correlation functions in eq.~(\ref{s corr func}), the general solution (\ref{v_sol}) allows us to write them all as
\begin{equation}\label{s corr func eig}
d_s(t,t') = \sum_{n=1}^3 e^{\lambda_n(t-t')} d_{s,n}(t'),
\end{equation}
with $d_{s,n}(t)=\left[M_n\vec{u}_{A(t)}\right]_3$ (the subscript denoting the third vector component), where $d_s$ denotes any of the correlation functions $\{d_s^+,\tilde{d}_s^+,d_s^-,\tilde{d}_s^-\}$, for which $A$ is taken, respectively, as $\{\delta a_s^{\dagger 2}\rho_s,\rho_s\delta a_s^{\dagger 2},\delta a_s^2\rho_s,\rho_s\delta a_s^2\}$. Let us emphasize that, just as with the pump mode, we have been able to write all the correlation functions of the signal mode into a form where the $t$ dependence only enters in simple exponential factors.

  

\cnb{Finally, let us show how this form for the correlation functions allows us to turn the c-MoP equations, which are coupled integro-differential equations, into coupled ordinary differential equations. For this aim, let us rewrite eqs.~(\ref{s c-MoP eqs}) and~(\ref{p c-MoP eqs}) as
\begin{align}
\dot{\rho}_s(t)=\mathcal{L}_s\rho_s(t)+\frac{\chi}{2}&\left[a_s^{\dagger 2}\Bra a_p\Ket(t)-a_s^2\Bra a_p\Ket\hspace{-0.06cm}^\ast\hspace{-0.06cm}(t),\rho_s(t)\right] 
\label{s c-MoP eqs TL}
\\
& +\left( \frac{\chi}{2}\right)^2\left\{\left[a_s^2\,,\,h_s(t)\right]+\text{H.c.}\right\},  \nonumber
\end{align}
\begin{align}
\dot{\rho}_p(t)=\mathcal{L}_p \rho_p&(t)+\frac{\chi}{2}\left[a_p \Bra a_s^2\Ket\hspace{-0.06cm}^\ast\hspace{-0.06cm}(t)-a_p^\dagger\Bra a_s^2\Ket(t)\,,\rho_p(t)\right] 
\label{p c-MoP eqs TL}
\\
& +\left(\frac{\chi}{2}\right)^2\left\{\left[a_p\,,\sum_{n=1}^3 h_{p,n}(t)\right]+\text{H.c.}\right\},  \nonumber
\end{align}
where we have defined the operators
\begin{equation}
\begin{split}
h_s(t) = \int_{0}^{t}dt^\prime e^{\mathcal{L}_{s}(t-t^\prime)}\mathcal{K}_s(t,t^\prime)\rho_s(t^\prime),
\\
h_{p,n}(t) = \int_{0}^{t}dt^\prime e^{\mathcal{L}_{p}(t-t^\prime)}\mathcal{K}_{p,n}(t,t^\prime)\rho_p(t^\prime),
\end{split}
\end{equation}
with the superoperator $\mathcal{K}_{p,n}$ defined as $\mathcal{K}_p$ in eq. (\ref{Kp}), but with the correlation functions $d_{s,n}(t)$ instead of $d_s(t)$. Using their definition, and the solutions found for the correlation functions, eqs. (\ref{p corr func}) and (\ref{s corr func eig}), their evolution equations are found to be}
%
%
%
\begin{align}
\partial _t h_s(t)&= (-\gamma_p +\mathcal L_s) h_s(t)  + \mathcal K _s(t,t)\rho_s(t), \label{h s eq}
\\
\partial _t h_{p,n}(t)&= (\lambda_n +\mathcal L_p) h_{p,n}(t)+\mathcal K_{p,n}(t,t)\rho_p(t). \label{h p eq}
\end{align}
%
%
%
%
%
%
\cnb{Together with eqs. (\ref{s c-MoP eqs TL}) and (\ref{p c-MoP eqs TL}), these form a closed set of coupled nonlinear ordinary differential equations for the reduced states $\rho_s$ and $\rho_p$, and the traceless operators $h_s$ and $\{h_{p,n}\}_{n=1,2,3}$. These are the equations that we analyze in the remainder of the paper.}

Overall we have shown for the example of the DOPO that it is indeed possible to rewrite the integro-differential c-MoP equations into a set of ordinary differential equations. The steps presented here are quite general and can be pursued for all c-MoP equations describing any physical system. The complexity of the resulting set of coupled equations will depend on the complexity of the subparts of the full quantum system, here given by the complexity of $\mathcal L_p$ and $\mathcal L_s$. 

Finally, we remark that the c-MoP equations preserve the trace and the hermiticity but they do not guarantee for the positivity of the density matrix. Such an issue is not unusual for projection operator theories, in fact, the same conditions can be found in the well established Redfield equations \cite{Redfield,Blum}. Obviously whenever the c-MoP equations provide a good approximation, they will yield a positive density matrix. Hence the positivity of the eigenvalues can be used as a consistency test for the approximation. 

%
%
\subsection{The adiabatic and the diabiatic limit}\label{sec:ada limit}
The two dissipation rates $\gamma_p$ and $\gamma_s$ set a time scale on which the pump and the signal, respectively, relax to the steady state of their unperturbed Liouvillians $\mathcal L_p$ and $\mathcal L_s$. In standard open system theory one relies on a separation of time scales between the system dynamics and the environment correlations. A similar reasoning is applied in adiabatic elimination approaches, where for the DOPO one relies on a time scale separation between signal and pump. The c-MoP theory can, in fact, be understood as a generalization of adiabatic elimination procedures where one has to consider the back-action of the ``system'' onto the ``environment''. We will now show that the effective equations for the reduced state of the signal known in the \emph{adiabatic} \cite{CarmichaelBook2} and the \emph{diabatic} \cite{FleischhauerMF} limit can, indeed, be obtained as limiting cases of the c-MoP equations.

The adiabatic limit in which the time scale of the pump mode is much faster than the time scale of the signal mode is defined such that $\gamma_p/\gamma_s\rightarrow \infty$ while $\gamma_s \gamma_p$ \cnb{is kept finite}. The diabatic limit describes the opposite scenario where $\gamma_p/\gamma_s\rightarrow 0$. \cnb{We proceed by comparing the Born terms with the free evolution operators $\mathcal L_p$ and $\mathcal L_s$, for which we consider the scaling of $h_s/\gamma_s$ and $h_{p,n}/\gamma_p$, which can be obtained by simple inspection of eqs. (\ref{h s eq}) and (\ref{h p eq}) divided by $\gamma_s$ and $\gamma_p$, respectively.
%
%
%

In the adiabatic limit, we infer from eq. (\ref{h p eq})/$\gamma_p$ that $h_{p,n}(t)/\gamma_p=0$ for all $n$ and $t\geq0$. Introducing this result into eq. (\ref{p c-MoP eqs TL}), we see that the state of the pump will be coherent with a field amplitude obeying the equation of motion
\begin{equation}\label{p field amplitude}
\partial_t \Bra a_p\Ket = \epsilon_p-\gamma_p(\Bra a_p\Ket+\tilde\alpha_p)-\frac{\chi}{2} \Bra a_s^2\Ket.
\end{equation}
On the other hand, eq. (\ref{h s eq})/$\gamma_s$ leads to $h_s(t)/\gamma_s = \mathcal{K}_s(t,t)/\gamma_s\gamma_p = \rho_s(t)\delta a_s^{\dagger2}(t)/\gamma_s\gamma_p$, where we have used eqs.~(\ref{Ks}) and~(\ref{p corr func}) and the fact that when the pump is in a coherent state all the expectation values in eq.(\ref{p corr func}) cancel. Introducing this result into eq. (\ref{s c-MoP eqs TL}), together with the steady-state solution of eq.~(\ref{p field amplitude}) for $\Bra a_p\Ket$, we end up with the effective master equation of the signal mode in the adiabatic limit
\begin{equation}\label{adiabatic eq}
\gamma_s^{-1}\partial_t \rho_s=\frac{\sigma}{2}\left[a_s^{\dagger\,2}-a_s^2\,,\,\rho_s \right]+\frac{g^2}{4} \mathcal D_{a_s^2}\rho_s+\mathcal D_{a_s}\rho_s,
\end{equation}
where $\sigma=\epsilon_p \chi/\gamma_p\gamma_s$ is an injection parameter corresponding to a coherent exchange of excitations between the signal and pump modes, while $g^2=\chi^2/\gamma_p\gamma_s$ accounts for signal photon pairs that are lost to the strongly damped pump mode.} Equation~(\ref{adiabatic eq}) has been extensively studied in the literature \cite{Wolinsky88,Drummond80,Drummond91}. It can be derived via standard adiabatic elimination which in the language of projection operator theory uses a time-independent projection superoperator $\mathcal P_{\text{ad}}$ projecting out the coherent laser field \cite{CarmichaelBook2}. Its action on the full density matrix is given by $\mathcal P_{\text{ad}}\, \rho(t)\equiv |\alpha\Ket \Bra \alpha| \otimes \rho_s(t)$, where $|\alpha\Ket$ is a coherent state with $\alpha=\epsilon_p/\gamma_p$. The fast exponential decay $e^{-\gamma_p (t-t')}$ of the pump correlation functions \cnb{allows in this case for a Markovian approximation in the Born terms, that is $\int_0^t dt' e^{\mathcal{L}_s(t-t^\prime)}\mathcal K_s(t,t')\rho_s(t')\approx\mathcal K_s(t,t)\rho_s(t)/\gamma_p$}.

\cnb{Let us now analyze the c-MoP equations in the diabatic limit. In this case, eq.~(\ref{h s eq})/$\gamma_s$ provides us with $h_s(t)/\gamma_s=0$, which introduced in eq. (\ref{s c-MoP eqs TL}) leads to an effective master equation}
\begin{equation}
\dot \rho_s(t) = \frac{\chi}{2} \left[ \Bra a_p\Ket a_s^{\dagger\,2}-\Bra a_p\Ket^* a_s^2\,,\rho_s(t)\right] +\mathcal D_{a_s} \rho_s(t) \,.
\end{equation}
for the signal state.
The pump state only enters this equation trough the amplitude $\Bra a_p\Ket$ which obeys eq.~(\ref{p field amplitude}) since $h_{p,n}$ is traceless.
\cnb{Noting that this equation is equivalent to eq.~(\ref{MF eqs}), we conclude that the diabatic limit reduces the full c-MoP equations to the mean-field equations}.
%

We emphasize that within these limits both eqs.~(\ref{MF eqs}) and~(\ref{adiabatic eq}) become exact. 
We have thus shown that the c-MoP theory provides us with exact equations of motion in the limits $\gamma_p/\gamma_s\rightarrow \infty$ (adiabatic) and $\gamma_p/\gamma_s\rightarrow 0$ (diabatic) where it therefore becomes equivalent with well established theories \cite{CarmichaelBook2,FleischhauerMF}. In the remainder of the paper we will step beyond these cases in which time-scale separation is present and use the c-MoP theory to \cnb{access the signal state in the $\gamma_p\approx\gamma_s$ scenario}.%
%
\begin{figure*}[tbp]
\centering
\includegraphics[width=\textwidth]{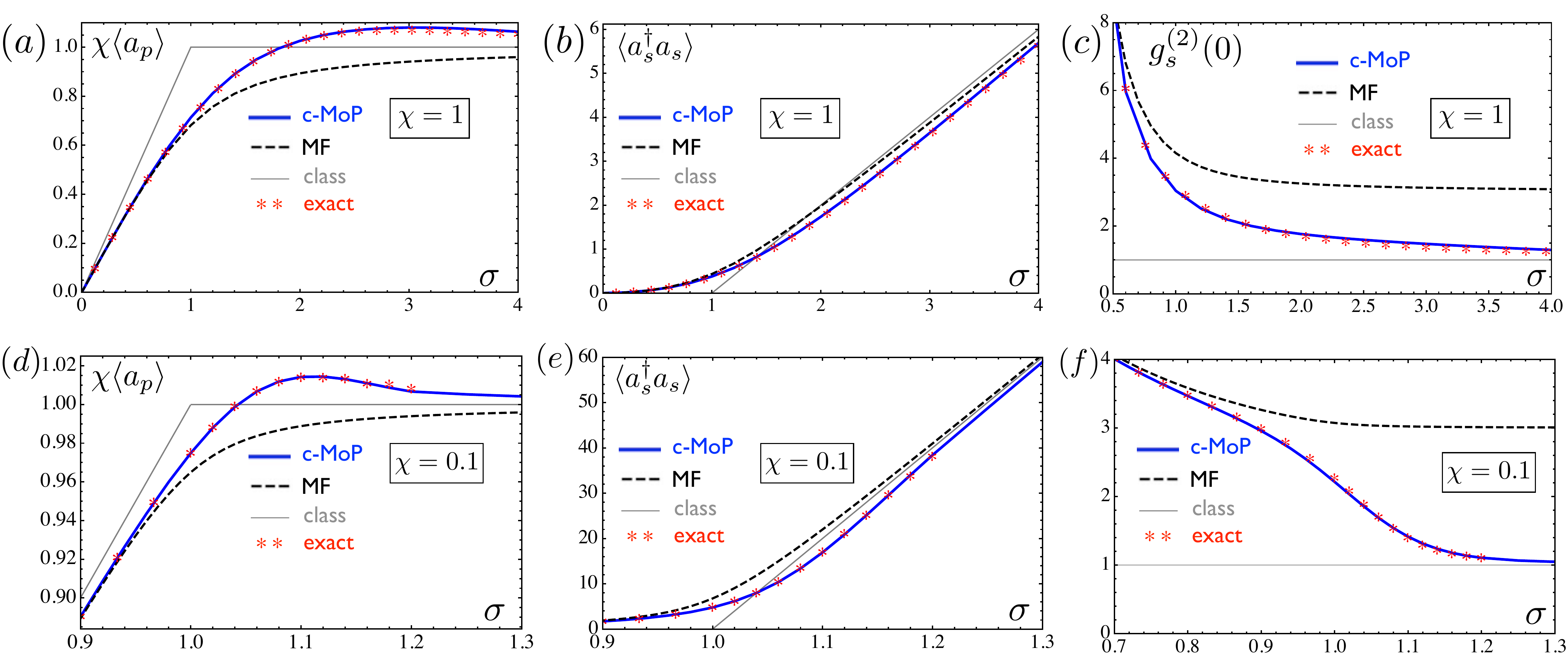}
\caption{\label{num c-MoP} Accuracy tests of the c-MoP theory for steady-state expectation values as a function of the injection parameter $\sigma$. In all plots we set $\gamma_p=\gamma_s=1$. The cases $\chi=1$ and $\chi=0.1$ are considered in $(a)$-$(c)$ and $(d)$-$(f)$, respectively. The rescaled pump amplitude $\chi \Bra a_p\Ket$ is shown in $(a)$ and $(d)$; $(b)$ and $(e)$ show the signal photon number $\Bra a_s^\dagger a_s\Ket$; finally, $(c)$ and $(f)$ show the $g^{(2)}$ function of the signal, which is equal to 1 for a coherent state (or a balanced mixture of them). The gray thin solid lines show the classical prediction from eqs.~(\ref{class eqs}), showing that the classical threshold where the signal field is switched on lies at $\sigma=1$. The blue solid curves represent the results obtained from the numerical solution of the c-MoP equations~(\ref{s c-MoP eqs TL}), (\ref{p c-MoP eqs TL}), (\ref{h s eq}), and (\ref{h p eq}). The red stars show the result obtained from the full master equation~(\ref{DOPO eqs}) up to injection parameters $\sigma$ where the numerics are tractable for us. Finally, the black dashed curves represent the mean-field theory, see eq.~(\ref{MF eqs}). Apart from the classical solution, all theories conserve the $Z_2$ symmetry, i.e. $\Bra a_s\Ket=0$.}
\end{figure*}
\section{Accuracy tests and full quantum states of the signal mode}\label{sec: full numerics}

In the previous section we have \cnb{shown how to deal with} the non-Markovian structure of the c-MoP equations. The only remaining difficulty is given by the quartic structure of the effective \cnb{equations of motion derived for the signal mode, eqs. (\ref{s c-MoP eqs TL}) and (\ref{h s eq})}. In this section we will treat the problem numerically in the \cnb{Fock} state basis by introducing a truncation $D_s$ for the Hilbertspace  $\mathcal H_s$ of the signal, \cnb{where $D_s$ is chosen such that the results for the observables we are interested in converge up to some desired accuracy}. Thus, the reduced state $\rho_s$ and the \cnb{operator} $h_s(t)$ will be $D_s\times D_s$ dimensional matrices. Instead of treating the pump mode in an analogous manner, we exploit the fact that the c-MoP \cnb{equations of the pump mode (\ref{p c-MoP eqs TL}) and (\ref{h p eq}) are quadratic in the bosonic operators}. As a consequence we are able to describe the pump state by a set of closed equations for five variables only, \cnb{the mode amplitude $\Bra a_p\Ket$ plus the fluctuations $\Bra a_p \delta a_p\Ket$ and $\Bra a_p^\dagger \delta a_p\Ket$ (note that the first two are complex variables)}. At the end, we are thus effectively left with two coupled differential equations for the matrices $\rho_s(t)$ and $h_s(t)$, \cnb{with the pump equations solved either in parallel numerically or analytically as a function of signal observables}.

In this section, we will compare the steady states of the classical theory from eq.~(\ref{class eqs}), the steady states of the mean-field equations~(\ref{MF eqs}), and the steady states of the c-MoP equations~(\ref{s c-MoP eqs}) and~(\ref{p c-MoP eqs}). In order to show the accuracy of the c-MoP equations we will also determine the steady state of the full Liouville-von Neumann equation~(\ref{DOPO eqs}) in parameter regimes where it is numerically tractable. \cnb{This numerical simulation is done as follows: first, we eliminate the large coherent background of the laser drive from the Liouvillian $\mathcal L$ by writing $a_p= \alpha_p+\delta a_p$, where $\alpha_p$ is taken to be the classical steady-state solution of eqs. (\ref{class eqs}); then, we use the superspace formalism, where the steady-state operator $\rho_{ss}$ and the Liouville superoperator $\mathcal{L}$ are represented, respectively, by a vector $\vec{\rho}_{ss}$ and a matrix $\mathbb{L}$, and $\vec{\rho}_{ss}$ can be found as the eigenvector with zero eigenvalue of $\mathbb{L}$ \cite{Drummond91,CarlosNumerics}. As the dimension of the matrix $\mathbb{L}$ is $(D_p\times D_s)^2$, with $D_p$ denoting the pump mode's Hilbert space dimension, this exact simulation is limited to small photon numbers.}

In all the simulations we consider cases without time-scale separation between the two modes and rescale all units to the dissipation rates, i.e. we put $\gamma_p=\gamma_s=1$. The only remaining parameters are the nonlinear coupling $\chi$ and the injection parameter $\sigma= \epsilon_p \chi$.

In Fig.~\ref{num c-MoP} we present results in parameter regimes where the full DOPO equation~(\ref{DOPO eqs}) can be solved numerically. In Figs.~\ref{num c-MoP}$(a)-(c)$ and \ref{num c-MoP}$(d)-(f)$ we show different steady-state observables for $\chi=1$ and $\chi=0.1$, respectively. It can be appreciated how the c-MoP results (blue solid line) coincide almost perfectly with the numerical results from the full master equation (red stars). The observables that we show are the pump mode's amplitude $\Bra a_p\Ket$ in Figs. \ref{num c-MoP}$(a)$ and \ref{num c-MoP}$(d)$, the signal photon number $\Bra a_s^\dagger a_s\Ket$ in Figs. \ref{num c-MoP}$(b)$ and \ref{num c-MoP}$(e)$, and the $g^{(2)}$ function $g_s^{(2)}(0)\equiv \Bra a_s^{\dagger\,2} a_s^2\Ket/\Bra a_s^\dagger a_s\Ket^2$ of the signal in Figs. \ref{num c-MoP}$(c)$ and \ref{num c-MoP}$(f)$. We also compare with the mean-field predictions of eqs.~(\ref{MF eqs}) (black dashed line), which in this context should be understood as the c-MoP theory up to first order, and with the classical steady-state solutions (gray thin solid line) given after eq.~(\ref{class eqs}). Let us remark that despite the nonlinear nature of the mean-field and the c-MoP equations, we only find one physical solution for each of them.

All four theories agree quite well far below the critical point $\sigma=1$ as the states of the signal and pump modes are close to vacuum and a coherent state induced by the external laser drive, respectively. Far above the threshold point, where the classical theory is expected to be approximately valid, we find that both the c-MoP predictions and the exact numerics agree well with the classical solutions for all observables, but with the fundamental difference that the classical theory breaks the $Z_2$ symmetry, while c-MoP and the exact solution preserve it. The mean-field solution, on the other hand, fails to describe the state of the signal above threshold as can be appreciated from the $g^{(2)}$ function in Figs. \ref{num c-MoP}$(c)$ and \ref{num c-MoP}$(f)$. As expected, mean-field theory and the classical theory break down in the vicinity of the threshold point.
Remarkably, this is not true for c-MoP which appears to give quasi exact results for all values of $\sigma$, even in cases where the interaction rate $\chi$ is comparable to all other system parameters.

%
\begin{figure*}[tbp]
\includegraphics[width=0.9\textwidth]{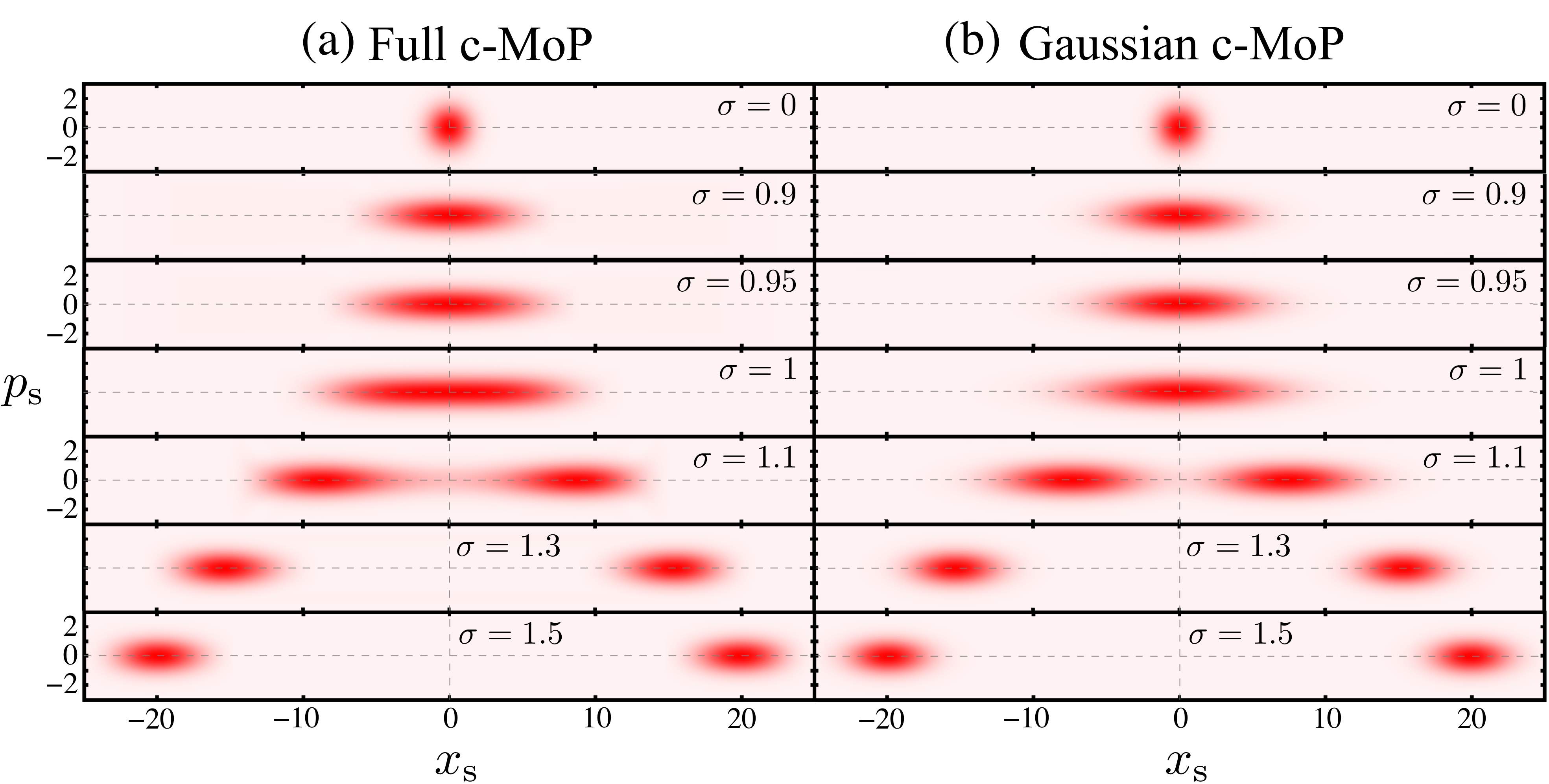}
\caption{\label{Wigner c-MoP} Wigner functions of the c-MoP density matrix for the signal mode without (a) and with (b) the Gaussian state approximation for $\gamma_p=\gamma_s=1$, $\chi=0.1$ and for different values of $\sigma$. \cnb{In the absence of injection, $\sigma=0$, the signal state is in vacuum}. Upon approaching the threshold, it becomes squeezed, with the highest squeezing levels obtained around $\sigma=1$. Above threshold two symmetric peaks appear and the squeezing \cnb{reaches some asymptotic value as we move away from threshold. Note how above threshold the state can be approximated by a balanced mixture of two symmetry breaking states. Indeed, let us remark that while for $\sigma<1$ we are plotting the unique solution that appears when applying the Gaussian state approximation onto the c-MoP equations (which we have called below threshold solution in the text), for $\sigma>1$ we have chosen to plot the Wigner function corresponding to a balanced mixture of the two above threshold symmetry breaking solutions with opposite phase which coexist with the non-symmetry breaking Gaussian solution}.}
\end{figure*}
\cnb{For the experimentally relevant scenario with $\chi\ll 1$, the Hilbert space dimension needs to be so large that we are not able to find the numerical solution of the full master equation~(\ref{DOPO eqs}) for injection parameters close to (or above) threshold. However, we can compare the c-MoP predictions (red stars), see Fig.~(\ref{GSA c-MoP}), with the perturbative approach which Drummond et al. (dark yellow dot-dashed line) developed in the vicinities of the critical point, by making a consistent multiple-scale expansion of the system's stochastic variables within the positive $P$ representation \cite{Kinsler95,Chaturvedi02}. This procedure has the virtue of being valid for any values of $\gamma_p$ and $\gamma_s$, and close to threshold, concretely for $|\sigma-1| < \chi/\sqrt{2\gamma_p\gamma_s}$, it is expected to be quasi-exact. As shown in Figs. \ref{GSA c-MoP}$(a)$ and \ref{GSA c-MoP}$(b)$, we find perfect agreement between this approach and the c-MoP theory for $\chi=0.01$.}

Overall, we have indeed shown the drastic impact of the Born terms, which do not only lead to a quantitative improvement as compared to the classical theory or to mean-field, but to a qualitatively different state of the signal mode. The classical theory predicts a coherent state, while the mean-field theory, i.e. the c-MoP theory up to first order, predicts a Gaussian state of the signal centered around $\Bra a_s\Ket =0$ \cite{Carlos14}. The c-MoP theory including the born terms, hence including signal-pump correlations within a projection operator based theory, is capable of finding the full quantum state of the signal which is neither coherent nor Gaussian as shown through the $g^{(2)}$ function in Figs. \ref{num c-MoP}$(c)$ and \ref{num c-MoP}$(f)$.

In order to illustrate the full quantum state, \cnb{we plot the Wigner function $W(x_s,p_s)$ of the signal density matrix obtained from the c-MoP equations in Fig.~\ref{Wigner c-MoP}$(a)$ for $\chi=0.1$ and different values of $\sigma$. Let us remark that in our case in which the Wigner function is positive everywhere in the phase-space formed by the quadratures $x_s=a_s^\dagger+a_s$ and $p_s=i(a_s^\dagger-a_s)$, it can be simply interpreted as the joint probability distribution describing the statistics of measurements of these observables \cite{Braunstein05,Weedbrock12,CarlosQI}. From a computational point of view, we evaluate it from the steady-state density matrix following the method detailed in \cite{Carlos14PRL}}. Far below threshold, the Wigner function shows a perfect vacuum for the signal state, see top panel of Fig.~\ref{Wigner c-MoP}$(a)$ for $\sigma=0$ as a reference. \cnb{As we cross through the critical point, two significant effects take place. First, approaching the threshold we find the well-known quadrature-noise reduction or \emph{squeezing} \cite{CarmichaelBook2,Eberle10,Mehmet08}, which is highest around the critical point $\sigma=1$ \cite{Chaturvedi02}, and reaches its asymptotic value $\Bra \delta p_s^2\Ket=(\gamma_s+\gamma_p)/(2\gamma_s+\gamma_p)$ for $\sigma\rightarrow\infty$, with corresponding \emph{antisqueezing} $\Bra \delta x_s^2\Ket=1+\gamma_s/\gamma_p$ \cite{Carlos14}. Second, as we cross the threshold we appreciate how the state develops two peaks centered (asymptotically) at the quadrature values predicted by the classical solution. Hence, even though the true quantum state never breaks the $Z_2$ symmetry, it does so in two qualitatively different ways depending on whether we are below or above threshold.}

In retrospect, \cnb{we see that the symmetry-breaking states predicted above threshold by the standard linearization approach correspond each to one of the} two distinct peaks appearing in the exact state. \cnb{Far above threshold} $\sigma\gg1$ the two peaks have zero overlap and \cnb{such states provide reasonable predictions} for all observables which are not sensitive to symmetry breaking, that is, all observables containing even numbers of signal field operators. \cnb{Of course, such a deficit can be corrected by simply using a balanced mixture of the symmetry-breaking states \cite{Carlos14}; this construction will guide us in the next section, where we will perform a Gaussian approximation which necessarily breaks the $Z_2$ symmetry}. It is then close to the critical point where both linearization and mean-field approaches fail, whereas c-MoP \cnb{provides an accurate description of the quantum state}.

Let us remark that we have compared the Wigner function obtained from the c-MoP theory with the reduced signal states obtained from the full master equation, which was only possible for $\sigma\leq1.2$, and found very good agreement, the differences being completely unnoticeable to the naked eye. We emphasize again that, with the numerical solution of the c-MoP equations we are able to find the full reduced density matrices of the modes away from the adiabatic limit. This is in contrast to other approaches such as stochastic simulations \cite{Kinsler95,Chaturvedi02,Chaturvedi99,Pope00} or the Keldysh formalism \cite{FleischhauerKeldysh,Mertens93,Swain93} which are naturally design to provide expectation values of the system operators. 

\section{Gaussian state Approximation within the c-MoP theory}\label{sec:GSA}

Despite the fact that the complexity of solving the c-MoP equations fully numerically scales in a more favorable way than the numerical complexity of the full master equation, 
it still requires to integrate a number of differential equations that scales quadratically with the dimension of the truncated Hilbert space for the signal field.
Therefore, it is very desirable to find an effective description of the underlying theory which is numerically more efficient and can thus cover the whole parameter space. In the remainder of this section, we implement such an idea by applying a \emph{Gaussian state approximation} (GSA) consistent with the c-MoP equations~(\ref{s c-MoP eqs}) and~(\ref{p c-MoP eqs}). 

\cnb{Another} great advantage of a Gaussian theory, apart from reaching the whole parameter space, is the efficiency in the evaluation of both steady states and dynamical quantities such as two-time correlation functions.
The disadvantage of a Gaussian theory, however, is the lack of quantitative accuracy especially in the vicinity of the critical point. \cnb{Nonetheless, as we show in the following, a Gaussian theory consistent with the c-MoP equations offers better quantitative accuracy than any of the previously developed Gaussian methods, particularly linearization around the classical solution or the recently-developed self-consistent linearization \cite{Carlos14}}.

The general procedure for finding a GSA for the state of a certain bosonic master equation is very simple. \cnb{In a first step, we write the bosonic operators as $a_j=\alpha_j+\delta a_j$,
with $\alpha_j=\Bra a_j\Ket$, such that $\Bra \delta a_j\Ket=0$. In the next step we find the evolution equation for the first and second moments, which will depend on higher-order moments in general. Thus, in the final step we assume the state to be Gaussian at all times, so that all higher order moments factorize into products of first and second order moments \cite{CarmichaelBook2,Carlos14}; in particular, we will encounter third order moments such as e.g. $\Bra \delta a_s^{\dagger 2} \delta a_s\Ket$ which vanish identically within the GSA, and forth order moments which factorize according to, e.g. 
\begin{equation}
\begin{split}
\Bra \delta a_s^{\dagger 4}\Ket&\approx 3 \Bra \delta a_s^{\dagger 2} \Ket^2, \\
\Bra \delta a_s^{\dagger 3}\delta a_s\Ket&\approx 3 \Bra \delta a_s^{\dagger 2} \Ket \Bra \delta a_s^{\dagger} \delta a_s \Ket, \\
\Bra \delta a_s^{\dagger 2}\delta a_s^2\Ket&\approx \Bra \delta a_s^{\dagger 2} \Ket \Bra \delta a_s^2 \Ket+2 \Bra \delta a_s^\dagger \delta a_s \Ket^2.
\end{split}
\end{equation}
After this final step, we are then left with a closed set of nonlinear equations for the amplitudes $\alpha_j$ and the second order moments of the fluctuations $\delta a_j$ that have to be solved self-consistently.} 

\begin{figure*}[tbp]
\centering
\includegraphics[width=\textwidth]{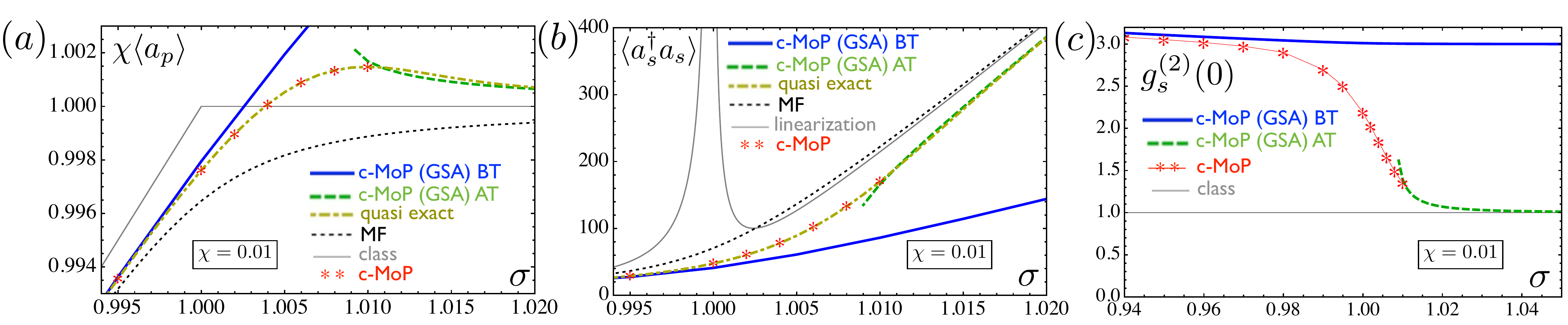}
\caption{\label{GSA c-MoP} Accuracy tests of the c-MoP theory with and without the Gaussian state approximation for steady-state observables as a function of the injection $\sigma$. In all plots we set $\gamma_p=\gamma_s=1$ and $\chi=0.01$. As in fig. \ref{num c-MoP}, we show the rescaled pump field amplitude $\chi \Bra a_p\Ket$, the signal photon number $\Bra a_s^\dagger a_s\Ket$, and the $g^{(2)}$ function of the signal, in $(a)$, $(b)$, and $(c)$, respectively. The red stars show the results obtained from the c-MoP eqs.~(\ref{s c-MoP eqs TL}), (\ref{p c-MoP eqs TL}), (\ref{h s eq}), and (\ref{h p eq}), up to injection parameters $\sigma$ where the numerics are tractable.
For comparison, the quasi exact method of Drummond and collaborators \cite{Kinsler95,Chaturvedi02} is shown as a dark yellow dot-dashed line. The blue solid and the green dashed curves represent the below and above threshold solutions, respectively, obtained from a Gaussian state approximation on the c-MoP equations. The black thin dotted curve displays the results of mean-field theory, see eq.~(\ref{MF eqs}), which in this context can be understood as the below threshold solution of a Gaussian state approximation on the full master equation (\ref{DOPO eqs}). Finally, the gray thin solid lines represent the prediction of the standard linearization theory in $(a)$ and $(b)$, and the coherent-state prediction $g^{(2)}=1$ of the classical equations~(\ref{class eqs}) in $(c)$.}
\end{figure*}

The standard linearization theory can be understood as a GSA on the full master equation, but with the exception that the amplitudes $\alpha_j$ are not determined self-consistently, but are obtained from the classical theory. As shown by the gray thin solid line in Fig.~\ref{GSA c-MoP}$(b)$ the complete \cnb{suppression of quantum fluctuations when determining these amplitudes} leads to unphysical results at the threshold point in the DOPO. 

The \emph{self-consistent linearization method}, as it is coined in Reference \cite{Carlos14}, goes one step beyond standard linearization by consistently finding \cnb{the amplitudes $\alpha_j$ from the GSA still applied to the full master equation. Due to the nonlinear nature of the resulting equations of motion one can find several solutions in a given point of parameter space. However, it was shown that at the end only two types of solutions were physical, qualitatively similar to the solutions found from standard linearization, but quantitatively regularized in such a way that the unphysical results of the latter disappear. In particular, a \emph{below threshold} (BT) solution was found, which does not break the $Z_2$ symmetry, i.e. $\alpha_s=0$, but in contrast to the classical theory exists for all values of the injection parameter, not only for $\sigma<1$. We also found two \emph{above threshold} (AT) solutions with opposite phase which break the symmetry, that is, $\Bra a_s\Ket=\pm|\alpha_s|\neq0$}, but appear only above a certain injection parameter $\sigma >1$ which is larger than the classical threshold value. Interestingly, we point out that the BT solution found through this self-consistent linearization is exactly equivalent to the mean-field theory introduced in Section \ref{sec:MF}.


Motivated by these findings, we apply a GSA to the c-MoP equations. Concretely, we calculate all first and second order moments of the pump and signal fluctuations from the c-MoP eqs.~(\ref{s c-MoP eqs TL}), (\ref{p c-MoP eqs TL}), (\ref{h s eq}), and (\ref{h p eq}), and apply the factorization of higher order moments as explained above. In strong contrast to the GSA on the full master equation, we do not need to assume a Gaussian state for the full state $\rho$ but only for the reduced state of the signal $\rho_s$. Hence, we expect similar qualitative results but with a higher quantitative accuracy.

Indeed, this is what we find and illustrate in Fig.~\ref{GSA c-MoP} for $\gamma_p=\gamma_s=1$ and $\chi=0.01$. We plot steady state expectation values for the pump amplitude $\chi \Bra a_p\Ket$ in Fig.~\ref{GSA c-MoP}$(a)$, the signal photon number $\Bra a_s^\dagger a_s\Ket= \Bra \delta a_s^\dagger \delta a_s\Ket+|\alpha_s|^2$ in Fig.~\ref{GSA c-MoP}$(b)$, and the $g^{(2)}$ function of the signal in Fig.~\ref{GSA c-MoP}$(c)$, all as a function of the injection parameter $\sigma$.
The blue solid line in Fig.~\ref{GSA c-MoP} shows the below threshold solution of the GSA on the c-MoP equation, while the green dashed line illustrates the above threshold solution. The latter fulfills $\Bra \delta a_s^\dagger \delta a_s\Ket\ll |\alpha_s|^2$ and is therefore \cnb{more likely to provide physically consistent results than the BT solution whenever they coexist}. In Fig.~\ref{GSA c-MoP}$(c)$ we show how the AT solution indeed gives the correct value for the $g^{(2)}$ function, \cnb{what indicates that each of the AT solutions corresponds to one of the peaks of the Wigner function, see Fig.~\ref{Wigner c-MoP}$(a)$, and considers Gaussian fluctuations around it. In order to illustrate this point even further, we show in Fig.~\ref{Wigner c-MoP}$(b)$ the Wigner function \cite{Braunstein05,Weedbrock12,CarlosQI} corresponding to the GSA on the c-MoP equations (as explained in the previous section, above threshold we take the balanced mixture of the two symmetry breaking solutions, such that the resulting state preserves the $Z_2$ symmetry).}

Importantly, there is an increased quantitative accuracy of the BT solution obtained from the c-MoP theory as compared with the mean-field theory (or the self-consistent linearization), see Figs.~\ref{GSA c-MoP}$(a)$ and \ref{GSA c-MoP}$(b)$, for parameters below and especially at the classical threshold point. As mentioned in Sec.~(\ref{sec: full numerics}) we test the accuracy of our method by comparing with the quasi exact method of Drummond and collaborators \cite{Kinsler95,Chaturvedi02}, illustrated by the dark yellow dot-dashed line in Figs.~\ref{GSA c-MoP}$(a)$ and \ref{GSA c-MoP}$(b)$. This increase in accuracy can be attributed to the born terms, since the mean-field equations can be understood on the one hand as the first order approximation of the c-MoP theory, and on the other hand as the below threshold solution of the GSA on the full master equation of the DOPO.

To summarize this section, we have shown that the c-MoP equations also provide a highly accurate Gaussian theory which is still as effective as every other linearized theory but, in contrast, it takes significant signal-pump correlations into account. \cnb{This is relevant because, as stated above, a Gaussian theory has the virtue that both steady-state as well as dynamical quantities such as two-time correlation functions can be found efficiently for any time and set of parameters}. To emphasize this practical aspect of the GSA, we will show in Sec.~\ref{sec:dyn} that the level of accuracy that we have found here in the evaluation of the steady states is also present in the transient time evolution.

\begin{figure*}[tbp]
\centering
\includegraphics[width=\textwidth]{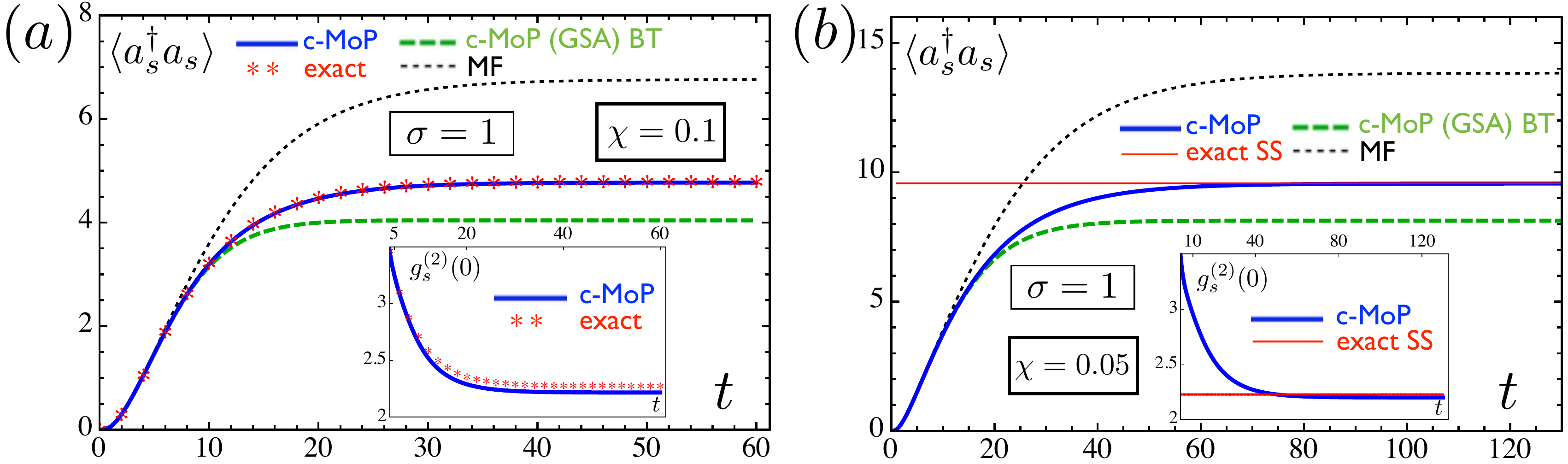}
\caption{\label{dynfig} Accuracy tests of the c-MoP theory for transient time evolution. The initial state is chosen to be the vacuum. We set $\gamma_p=\gamma_s=1$, investigate the classical threshold point $\sigma=1$, and choose $\chi=0.1$ in $(a)$ and $\chi=0.05$ in $(b)$. The plots display the signal photon number as a function of time in units of the dissipation rates, while the insets show the $g^{(2)}$ function of the signal mode. The red stars in $(a)$ show the result obtained from the numerical simulation of the full master equation~(\ref{DOPO eqs}), while the red line in $(b)$ indicates its steady-state values only. The blue solid curves represent the results obtained from the numerical integration of the c-MoP eqs.~(\ref{s c-MoP eqs TL}), (\ref{p c-MoP eqs TL}), (\ref{h s eq}), and (\ref{h p eq}). Finally, the green dashed and black dotted lines represent the time evolution obtained from a Gaussian state approximation on the c-MoP equations and the full master equation (mean-field), respectively.}
\end{figure*}

\section{Dynamics} \label{sec:dyn}

So far we have only presented steady state quantities for the various methods of our interest. In this section we will briefly elaborate on the possibility to simulate dynamical evolution as well. The steady state of the full master equation~(\ref{DOPO eqs}) can be understood as an eigenvector corresponding to the zero eigenvalue of the Liouvillian
$\mathcal L$ such that $\mathcal L \rho_{ss}=0$. The formal solution for the time evolving state which can be written as $\rho(t)=e^{\mathcal L t} \rho(0)$, on the other hand, involves all eigenvalues of the Liouvillian. Hence, it is a priori not clear whether \cnb{a given approximate method used for the evaluation of the steady state of $\mathcal L$ will provide the same degree of accuracy when used for transient time evolution.} 

In order to investigate this open issue we simulate the time dynamics of the various approximate methods that we have introduced, and compare their results with an exact simulation of the full master equation~(\ref{DOPO eqs}) in regions of the parameter space where it is numerically tractable. We analyze a situation in which the input laser drives the system from the initial vacuum to its steady state. Figure~\ref{dynfig} shows the signal photon number as a function of time at the classical threshold point $\sigma=1$, for $\gamma_p=\gamma_s=1$, and for $\chi=0.1$ in Figure~\ref{dynfig}$(a)$ and $\chi=0.05$ in Figure~\ref{dynfig}$(b)$. 
The red stars in Fig.~\ref{dynfig}$(a)$ illustrate the result obtained from the numerical simulation of the full master equation~(\ref{DOPO eqs}), while the red line in Fig.~\ref{dynfig}$(b)$ illustrates the steady state value of the observables only, \cnb{since the small value of $\chi$ prevented us from being able to simulate the whole dynamics in this case. On the other hand,} the blue solid curves represent the results obtained from the numerical integration of the c-MoP equations \cnb{as explained in Section \ref{sec: full numerics}. Finally,} the green dashed and black dotted lines represent the time evolution obtained from a \cnb{GSA on} the c-MoP equations and the full master equation, respectively.  

\cnb{Remarkably, Fig.~\ref{dynfig}$(a)$ shows that the level of accuracy found dynamically for the various approximations is similar to the ones that we already encountered when evaluating steady-state quantities. In particular, it is apparent that, at any point in time, the GSA on the full master equation is less accurate than the GSA on the c-MoP equations, which in turn does not have the remarkable level of accuracy shown by the full c-MoP numerical simulation, which almost coincides with the numerics of the full master equation at all times. It is important to note that the evolution of the $g^{(2)}$ function shown in the inset of Fig.~\ref{dynfig}$(a)$ suggests that, indeed}, the c-MoP equations are able to map the full quantum state of the signal in the course of time.

A numerical simulation for the parameter set chosen in Fig.~\ref{dynfig}$(b)$ demands minimal Hilbert space dimensions of $\dim \mathcal H_p= 6$ and $\dim \mathcal H_s= 120$ in order to reach convergence up to an accuracy of $10^{-2}$ for the relevant observables. Thus, \cnb{while the c-MoP approach requires a simulation of a set of $28\,811$ coupled nonlinear differential equations, in the case of the full master equation one has to integrate $518\,400$ coupled linear differential equations, which has precluded us from being able to simulate the dynamics from it}. Therefore we only show steady-state observables of the full master equation for these case. 

\cnb{Figs.~\ref{dynfig}$(a)$ and \ref{dynfig}$(b)$ further illustrate the scaling of various quantities with the nonlinear coupling $\chi$ at the critical point. In particular, note how both the signal photon number and the time that the system needs to reach the steady state double when $\chi$ is reduced by half. The latter is known in the literature as \emph{critical slowing down} \cite{Kinsler95}, and just as the signal photon number, it was predicted to scale with $\chi^{-1}$ \cite{Wolinsky88,Kinsler95,FleischhauerKeldysh}, in agreement with our c-MoP simulation. Hence, we can appreciate the practical use of a Gaussian theory by considering that to simulate an experimentally relevant scenario where $\chi\ll 1$, dynamical quantities would require extremely long simulation times, which, as explained before, can be efficiently handled with a GSA on the c-MoP theory, but not by its full numerical simulation. As an example, we have checked that for $\chi=0.01$ a GSA on the c-MoP equations requires a normalized time of approximately $300$ to reach the steady state, again in agreement with the $\chi^{-1}$ scaling, as it can be appreciated in Fig. \ref{dynfig}$(a)$ that such time is about 10 times smaller for $\chi=0.1$.}

\section{conclusions and outlook} \label{conclusions}
In conclusion we have exemplified the applicability of the self-consistent projection operator theory to nonlinear quantum optical systems on the case study of the degenerate optical parametric oscillator. Our theory generalizes mean-field approaches and in particular adiabatic elimination methods to settings without time-scale separation. The effective master equations can be solved efficiently despite their non-Markovian structure. We have demonstrated the high degree of accuracy of our method and revealed its capability to determine the exact quantum states below, at, and above the classical threshold for both stationary states and dynamical evolution.  

In addition, we developed a linearized theory consistent with the self-consistent Mori projector equations and showed its accuracy far beyond any known linearized theories. We expect our Gaussian method to be particularly useful in the context of hybrid systems such as optomechanical parametric oscillators \cite{OMPOCarlos,Nori}, \cnb{where fields of quantum nature with no coherent background are coupled to mechanical elements}. Some intriguing tasks for future research would include applying the c-MoP approach to investigate dynamical questions, e.g. investigate tunneling times between the two symmetry breaking states in parameter regimes away from the adiabatic limit \cite{Drummond91}, simulate quantum quenches in a driven-dissipative scenario \cite{Keeling}, and investigate the effect of small symmetry breaking perturbations on both the dynamics and steady states.


\acknowledgments
The authors thank Mehdi Abdi, Johannes Lang, Tao Shi, Yue Chang, Eugenio Rold\'an, Francesco Piazza, and Peter D. Drummond for fruitful discussions and comments. This work has been supported by the German Research Foundation (DFG) via the CRC 631 and the grant 
HA 5593/3-1. CN-B acknowledges funding from the Alexander von Humbolt Foundation through their Fellowship for Postdoctoral Researchers.
%
%
\appendix

\end{document}